# Fermi level influence on the adsorption at semiconductor surfaces – ab initio simulations

StanisławKrukowski*[1,2], Paweł Kempisty[1], Paweł Strąk[1]

[1]Institute of High Pressure Physics, Polish Academy of Sciences, Sokołowska 29/37, 01-142 Warsaw, Poland

[2]Interdisciplinary Centre for Materials Modelling, Warsaw University, Pawińskiego 5a, 02-106 Warsaw, Poland

Chemical adsorption of the species at semiconductor surfaces is analyzed showing the existence of the two contributions to adsorption energy: bond creation and charge transfer. It is shown that the energy of quantum surface states is affected by the electric field at the surface, nevertheless the potential contribution of electron and nuclei cancels out. The charge transfer contribution is Fermi level independent for pinned surfaces. Thus for Fermi level pinned at the surface, the adsorption energy is independent on the Fermi energy i.e. the doping in the bulk. The DFT simulations of adsorption of hydrogen at clean GaN(0001) and silicon at SiC(0001) surfaces confirmed independence of adsorption energy on the doping in the bulk. For the Fermi level nonpinned surfaces the charge contribution depends on the position of Fermi level in the bulk. Thus adsorption energy is sensitive to change of the Fermi energy in the bulk, i.e. the doping. The DFT simulations of adsorption of atomic hydrogen at 0.75 ML hydrogen covered GaN(0001) surface confirmed that the adsorption energy may be changed by about 2 eV by the doping change from n- to p-type.

## 1. Introduction

Surfaces attracted a considerable interest of a large number of researchers conducting investigations of their equilibrium and kinetic properties that constituted an important part of solid state physics and chemistry, bringing a number of important discoveries. In addition to the considerable importance of basic research, the molecular processes determining physical properties of the grown crystals and epitaxial layers have immense technological importance.

Among the solid surfaces these of metals and semiconductors constitute the two most prominent categories. They differ principally by different bonding, being directly isotropically cohesive in the case of metals and directional in the case of semiconductors. Thus the metal surfaces attach active species indiscriminately, remaining structurally relatively weakly affected by the adsorption processes.[1,2] Differently, the semiconductor surfaces are prone to reconstruction which may be drastically changed by the presence of adsorbed species.[3,4]

In addition to the above mentioned chemical and structural properties, the electric features of the two types of the solids and their surfaces are utmost diverse. The metals are relatively immune to any influence with the Fermi level set and the screening lengths of fraction of Angstroms. Thus their surfaces are directly simulated by *ab intio* calculations, using either slab of cluster models with a great success.[5]

The electric properties of semiconductor bulk and surfaces are far more susceptible to manipulation. By appropriate doping, the Fermi level may be shifted across the bandgap, to achieve, in some cases degenerate gas in the conduction or valence bands. Thus, the Fermi energy may be easily shifted by several electronvolts in the case of wide bandgap compounds. In addition to that, the semiconductor surfaces may be electrically charged, shifting the energy of states in the vicinity of surfaces. Typically, these charges are screened by the mobile or localized charges, over the screening distances of the range of extending from few Angstroms to microns. Finally, the Fermi level maybe free or pinned by the surface states which entails charge transfer to or from the surface and appropriate band bending to adjust to the Fermi energy in the bulk.[3,4]

It is therefore natural to expect that the simulation procedures of the semiconductor surfaces should be more sophisticated than these used for metals. Until recently the state of art was fairly disappointing. Usually, the slab models were employed with the opposite side surface of the solid body cut off, creating broken bonds which are saturated by integer or fractional charge hydrogen termination atoms, usually satisfying the electron counting (EC) rule. The device, essentially based on simple tight binding arguments was considered to be satisfactory that was supported by the claims that the system properties were size independent.[6-8]

In recent years a more sophisticated approach was proposed, based on the observation that by proper manipulation of the surface termination atoms by change of their location or their charge or both, the average electric field within the slab may be changed.[9-12] The model was applied in the studies of GaN(0001)[9-13] and SiC(0001) and SiC(000$\underline{1}$) polar surfaces.[14] Naturally, the change of the energies of the quantum states by the field at the surface was denoted as the surface states Stark effect SSSE.[11-14] In addition to the surface termination manipulation, an alternative approach was proposed in the recent edition of SIESTA shareware. The second approach, in principle, could be also applied to surface charged states.[15]

In the present study, we analyze the influence of the position of Fermi level on the adsorption energy of different species. Using GaN(0001) surface, and appropriate slab models, the role of Fermi level in the bulk will be investigated for both pinned and nonpinned cases, employing various adsorbing species. The role of the surface charge type, donor or acceptor and its change during the process will be elucidated. It will be shown that metal surface analogy is not well suited for the description of the semiconductor surface. The different behavior of the semiconducting surfaces will be discussed.

**2. The simulation procedure**

In part of the calculations reported below, a freely accessible DFT code SIESTA, combining norm conserving pseudopotentials with the local basis functions, was employed.[16-18] The basis functions in SIESTA are numeric atomic orbitals, having finite size support which is determined by the user. The pseudopotentials for Ga, H and N atoms were generated, using ATOM program for all-electron calculations. SIESTA employs the norm-conserving Troullier-Martins pseudopotential, in the Kleinmann-Bylander factorized form.[19] Gallium 3d electrons were included in the valence electron set explicitly. The following atomic basis sets were used in GGA calculations: Ga (bulk) - 4s: DZ (double zeta), 4p: DZ, 3d: SZ (single zeta), 4d: SZ; Ga (surface)- 4s: TZ (triple zeta), 4p: TZ, 3d: SZ, 4d: SZ; N (bulk) - 2s: TZ, 2p: DZ;  N (surface)- 2s: TZ, 2p: TZ, 3d: SZ; H - 1s: QZ (quadruple zeta), 2p: SZ and H (termination atoms)  1s: DZ, 2p: DZ, 3d: SZ. The following values for the lattice constants of bulk GaN were obtained in GGA-WC calculations (as exchange-correlation functional Wu-Cohen (WC) modification of Perdew, Burke and Ernzerhof (PBE) functional[20,21]: a = b = 3.2021 Å , c = 5.2124 Å. These values are in a good agreement with the experimental data for GaN: a = 3.189 Å and c = 5.185 Å [22]. All the presented dispersion relations are plotted as obtained from DFT calculations, burdened by a standard DFT error in the recovery of GaN bandgap. In the present parameterization, the effective bandgap for bulk GaN was 1.867 eV.

In the case of the slab, the gap is additionally affected by localization in finite thickness increasing the gap to the following values: 20, 10 and 8 GaN layers: 1.925 eV, 2.123eV and 2.228 eV, respectively. Hence, in order to obtain a quantitative agreement with the experimentally measured values, all the calculated DFT energies that were obtained for 10 GaN layers slabs, should be rescaled by an approximate factor $\alpha = E_{g\text{-exp}}/E_{g\text{-DFT}}=3.4eV/2.13eV \approx 5/3 \approx 1.6$. Integrals in k-space were performed using a 3x3x1 Monkhorst-Pack grid for the slab with a lateral size 2x2 unit cell and only Γ-point for 4x4 slabs.[23] As a convergence criterion, terminating a SCF loop, the maximum difference between the output and the input of each element of the density matrix was employed being equal or smaller than $10^{-4}$. Relaxation of the atomic position is terminated when the forces acting on the atoms become smaller than 0.02 eV/Å.

In a simulation of SiC(0001) surface, a commercially available VASP DFT code, developed at the Institut für Materialphysik of Universität Wien was also used.[24-26] Optimization of ionic positions was performed using Generalized Gradient Approximation (GGA) energy functional in order to obtain properly relaxed structures. In the first instance, a standard plane wave functional basis set, as implemented in VASP with the energy cutoff of 29.40 Ry (400.0 eV), was adopted.[27] The Monkhorst-Pack grid: (3x3x3), was used for k-space integration.[23] For Ga and N atoms, the Projector-Augmented Wave (PAW) potentials for Perdew, Burke and Ernzerhof (PBE) exchange-correlation functional, was used in Generalized Gradient Approximation (GGA) calculations.[20] The energy error for the termination of electronic self-consistent (SCF) loop was set equal to $10^{-6}$. For the exchange-correlation functional, generalized gradient approximation (GGA) in Perdew, Burke and Ernzerhof (PBE) approximation was used.[20] The plane wave basis cutoff energy was set to 400 eV. These parameters recover basic structural and energetic properties of 2H SiC with good accuracy.[14] The lattice parameters of 2H SiC obtained from the DFT calculations were: a = 3.092 Å and c = 5.074, compared well to the experimental data a = 3.079 Å and c = 5.053 Å.[28,29]

### 3. The Fermi level pinned surfaces

The analysis of the adsorption energy is conveyed using the absolute energy scale in which the energy of the valence and conduction states in the bulk and the separate atom/molecule quantum states have well prescribed values as shown in Fig. 1. In a vast majority of the atomic configurations, the Fermi level is pinned at the semiconductor surface that entails appropriate change of the charge on the surface state thus adjusting to the change of Fermi level in the bulk. Inadvertently, that entails emergence of the electric field, in accordance with the bands alignment in the bulk and at the surface. In spite of this, the energy of surface state pinning Fermi level is determined by the position of Fermi level in the bulk.

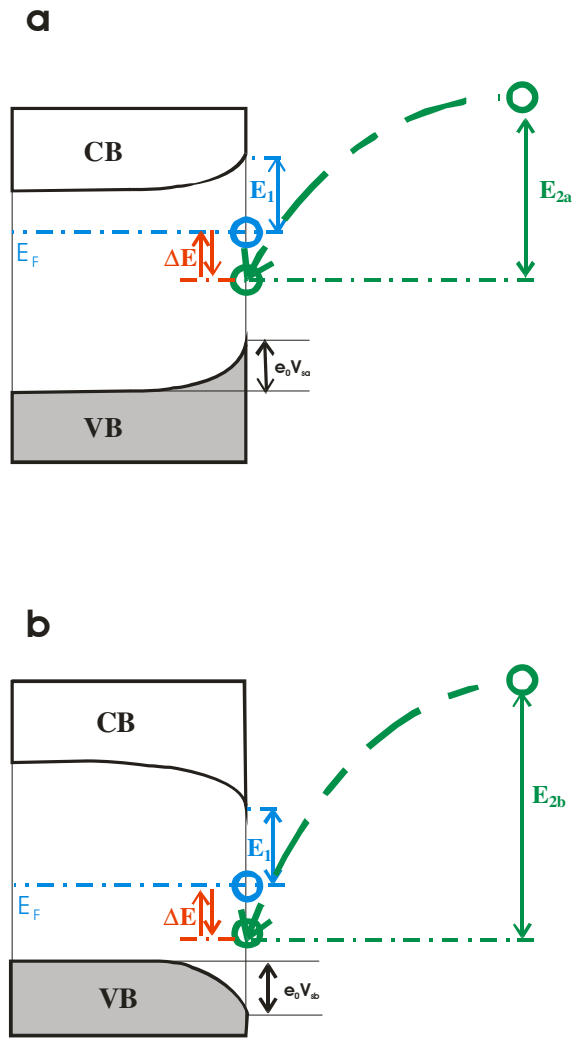

Fig. 1. The energy of the quantum states in the absolute scale, changed during adsorption process at surfaces with Fermi level pinned by: a - surface acceptor, b - surface donor. The change of the energy of electron quantum state of the adsorbing species are denoted by green color. Additional contribution to the adsorption energy brings the charge transfer between the surface states and the bulk. The electrons are transferred from or to the states in the bulk, of the energy close to Fermi level that is denoted by red color. The energy $E_1$ difference between the pinning surface state and the conduction band minimum is denoted by blue color.

Thus various doping in the bulk with the Fermi level pinning leads to emergence of the electric fields at the surface of different direction that are simulated using slab termination in the way recently invented for simulation of the fields at the surface[9-14]. By the change of occupation, the type of surface state, both donor and acceptor states could be simulated, enforcing the upward and downward band bending, as shown in Fig. 1 and b, respectively.

Adsorption of the molecule/atom leads to the change of the energy of quantum states of the adsorbed species. In Fig. 1 the molecule/atom quantum states energy evolution, from far distance to the adsorbed position, is presented. The depicted surface has the quantum state that pins Fermi level at the surface. The pinning state has higher energy for n- than for p-type semiconductor in the absolute scale. Naturally the energy of the surface state created by the attached molecule/atom need not to be equal to the energy of pinning state but it differs by the same difference $\Delta E$, equally for both p- and n-type bulk.

Additionally, as shown recently, the energy $\Delta E$ is only slightly affected by the reverse of the field direction, the effect that could be neglected in the first approximation. Naturally, the energy of the molecular/atomic states in initial, far distance configuration is identical in the absolute scale. The energy of these states in the final position, for the species adsorbed at the surface is different, on the virtue of pinning it is determined by the Fermi energy in the bulk and the surface state energy difference $\Delta E$. Therefore the adsorption caused change of the energy of molecular/atomic quantum states is different for n and p-type.

Since the difference of the energy of pinning state and the bands at the surface is different form the difference of the Fermi level and the bands in the bulk, the bands are bent upward (surface acceptor) or downward (surface donor), due to the surface state charge related electrostatic effect. It has to be accounted that the energy change of the negatively charged electron state is equal in magnitude but of the opposite sign to the energy change of the positively charged nuclei. Thus these two contributions cancel. In the result the adsorption energy in the process at the surface with Fermi level pinned should be identical provided the surface-bulk charge transfer vanish.

In addition to the above described zero charge transfer process, the adsorption is frequently accompanied by the electron transfer to or from the bulk, affecting the charge state of the surface. The surface charge state may be determined from the average potential slope as it is shown below. Naturally the electron is transferred to or from the Fermi level (more precisely, to or from the defect state of the energy very close to the Fermi energy). As the Fermi level is pinned by the surface state, the accompanying energy change may be calculated as the energy difference $\Delta E$ of the pinning state and the state created by molecule/atom adsorption. In the case when this is the same state, the contribution is zero while in the case of the different two states, the contribution is finite but it is identical for p- and n-type semiconductors. Thus the adsorption energy should be identical for the Fermi level pinned at the n- and p-type semiconductor surfaces, i.e. it is independent of the doping in the bulk.

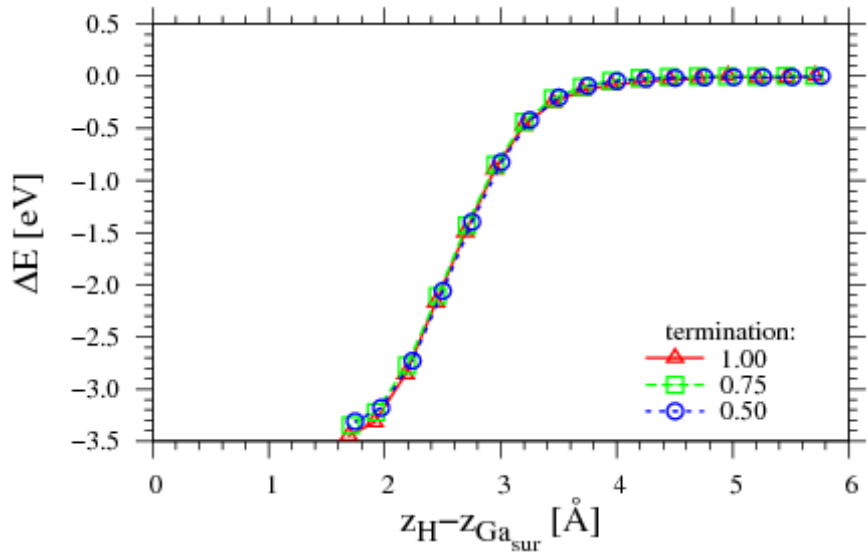

Fig. 2. Energy of the system during adsorption of single hydrogen atom on 8 double atomic layers thick 4 x 4 GaN slab, terminated by pseudo-hydrogen atoms with atomic number Z = 0.50, Z = 1.00 and Z = 0.75, representing surface donor, acceptor and neutral clean GaN(0001) surface (SIESTA).

As it is shown in Fig. 2, the adsorption energy of hydrogen atom, determined from DFT calculations, does not depend drastically on the field at the slab (or equivalently on the excess electric charge on the surface state). This fairly standard result may be illustrated by the band diagram extracted from the atom projected density of states (PDOS), presented in Fig. 3. Fermi level is pinned at the surface by the surface state emerging from the top gallium atom broken bonds.

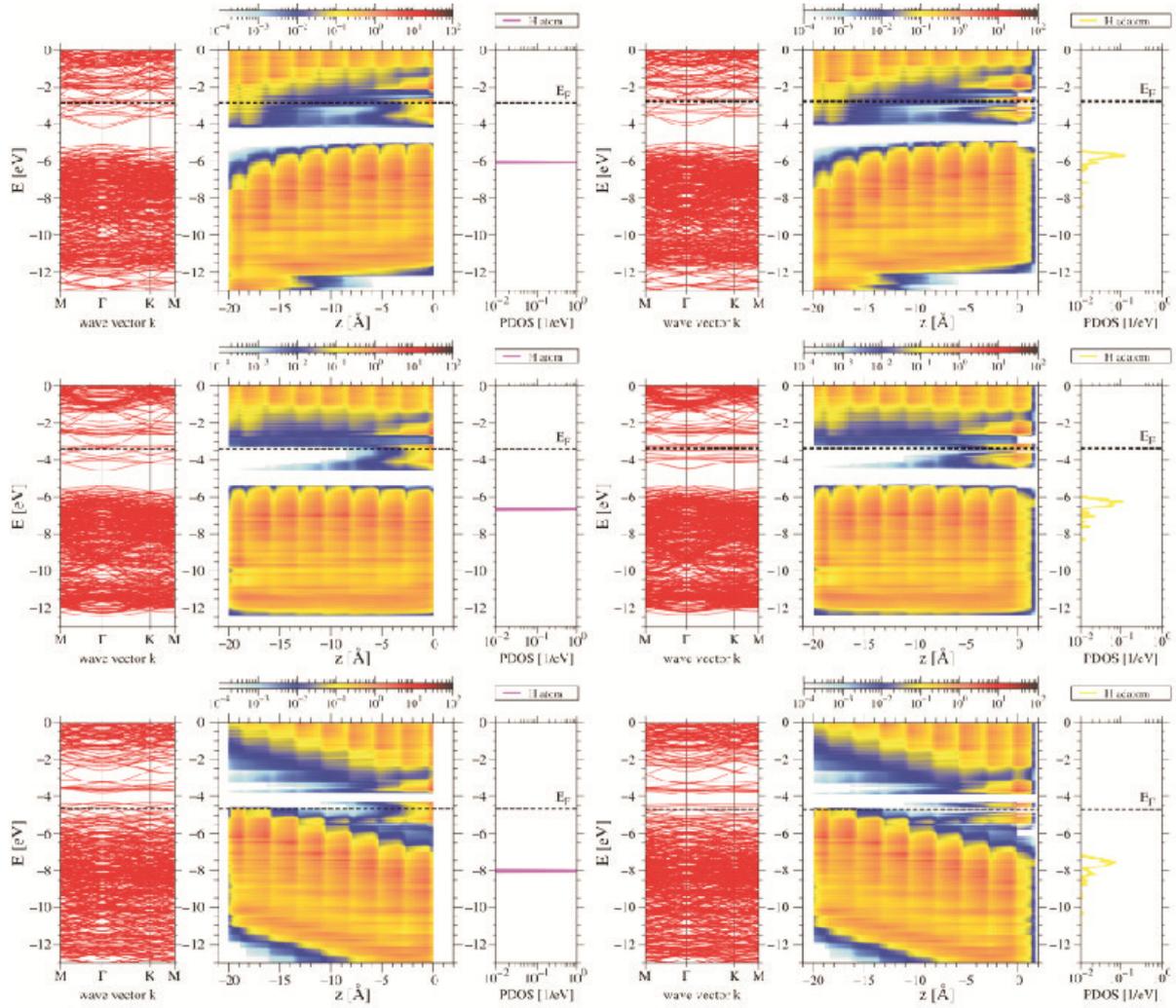

Fig. 3. Band diagram, atom row projected density of states and density of states projected on hydrogen atom, of the 4 x 4 slab before (left) and after (right) adsorption of single hydrogen atom at the clean GaN(0001) surface. The three cases present surface acceptor, neutral and surface donor from the top to the bottom, respectively.

These conclusions are illustrated by detailed study of adsorption of hydrogen at clean GaN(0001) surface. The process was investigated recently using 2 x 2 and 4 x 4 slabs, proving that the hydrogen atom is strongly attached at the surface, independently of the direction of the field.[13] The adsorption the energy of single hydrogen atom descending on the surface of 4 x 4 GaN slab representing GaN(0001) surface was high, equal to 3.45, 3.35 and 3.30 eV, for surface donor, neutral surface and surface acceptor, respectively. It is worth to note that the slope of the band diagram is changed in the process, indicating the considerable charge transfer between the surface and the bulk. Nevertheless, in accordance to the above arguments, the adsorption energy is approximately equal for all cases.

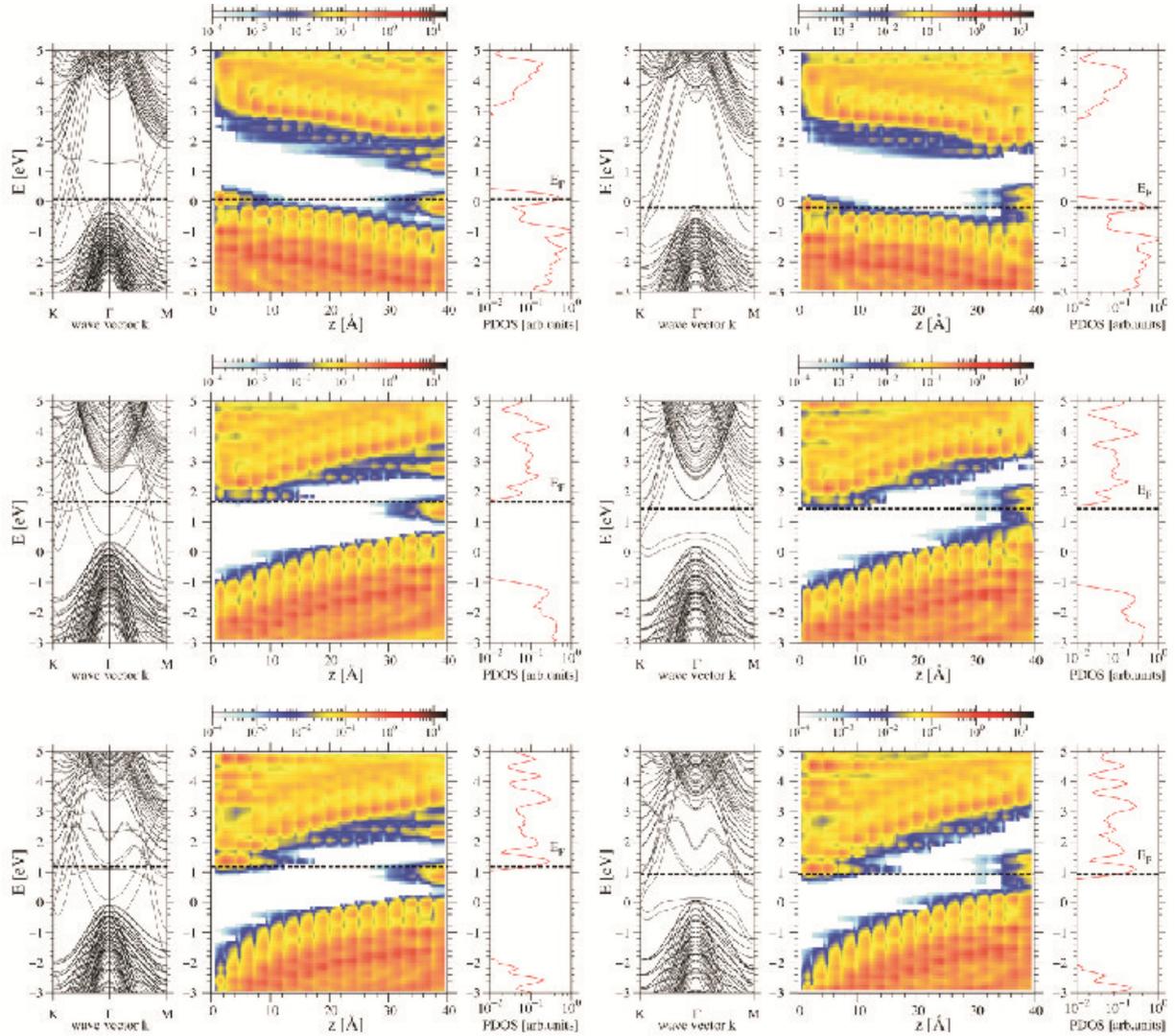

Fig. 4. Band diagram, the atom row projected density of states (spatial variation of the bands) and density of states projected on hydrogen atom, of the 1 x 1 slab before (left) and after (right) adsorption of single silicon atom at the clean SiC(0001) surface. The three cases present surface donor, neutral and surface acceptor from the top to the bottom.

The diagram present also the energy of hydrogen quantum states, separate and attached at the surface. As it is shown, the energy is changed, due to bonding to the surface. The state is change from zero dispersion molecular state to wide dispersion surface state, thus the bond creation contribution is proved to exist.

In order to verify these conclusion for different surfaces, the adsorption of single Si atom on SiC(0001) surface was considered using 1 x 1 eight Si-C double layer thick slab. As in the previous case, the Fermi level is pinned at the surface, as shown in Fig. 4. The surface adsorption energies obtained for surface donor, neutral and surface acceptor were equal to 2.02, 2.03 and 2.25 eV respectively. Thus the adsorption energy variation is of order of 0.2 eV, relatively small fraction of the total adsorption energy. From the band diagrams it follows that the charge transfer is considerable. Nevertheless, this charge transfer is accompanied by the small variation of the adsorption energy for these cases again confirming the above described conclusions.

Another important issue to be solved is the dependence of the adsorption energy on the height of Fermi level, i.e. doping in the bulk. In Figs. 5 and 6, the hydrogen adsorption on the surface of differently doped GaN bulk is investigated. The Fermi level is changed by the addition of Si or Mg atoms to the slab so that the obtained results correspond to semi-insulating, n- or p-type material. In standard DFT simulations for Ga-N slab, the position of Fermi level is not controlled directly, which usually corresponds to semi-insulating crystal.

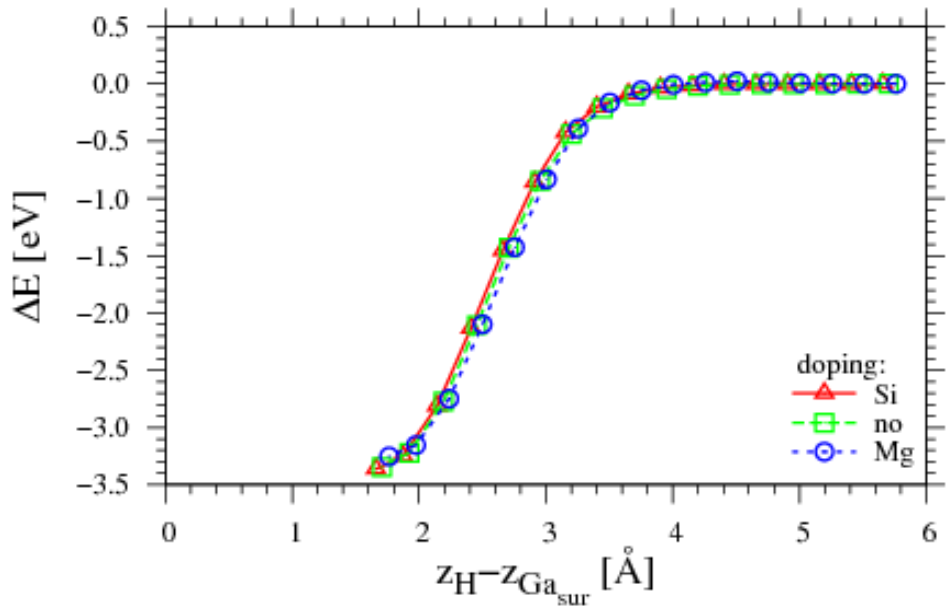

Fig. 5. Energy of the system during adsorption of single hydrogen atom on 8 double atomic layers thick 4 x 4 GaN slab representing clean GaN(0001) surface: no - GaN slab for semi-insulating bulk, Si - two Ga atoms replaced by Si atoms, representing n - GaN, Mg - two Ga atoms replaced by Mg atoms - for p - GaN (SIESTA).

As it is shown in Fig. 5, the adsorption energy does not depend on the doping within the slab. The doping directly affects the position of Fermi level as it is depicted in Fig. 6 where the band diagram, space projected and atom DOS are shown.

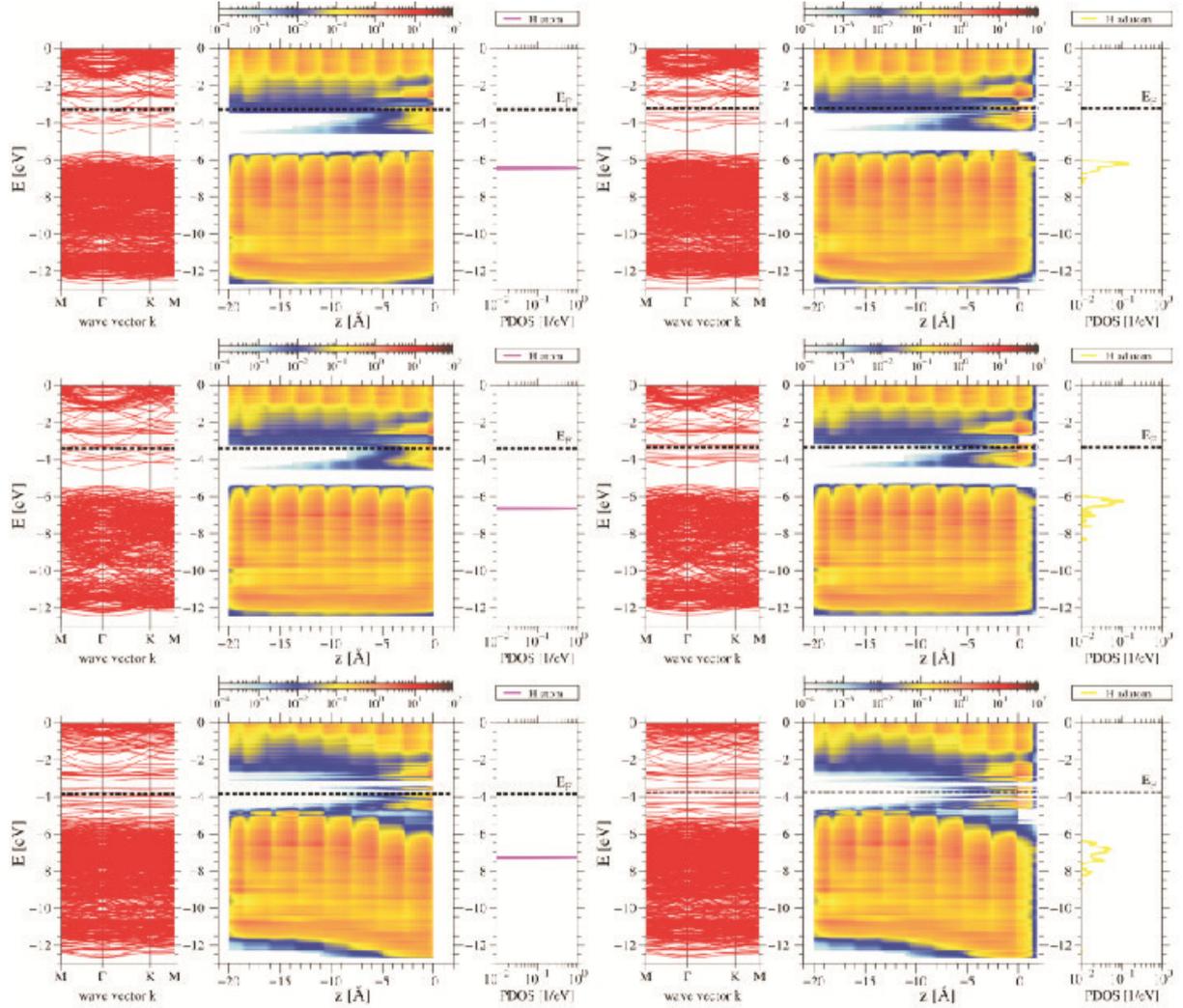

Fig. 6. Band diagram, the atom row projected density of states (spatial variation of the bands) and density of states projected on hydrogen atom, of the 4 x 4 slab before (left) and after (right) adsorption of single hydrogen atom at the clean GaN(0001) surface. The three cases present n-type (Si doped), neutral (undoped) and p-type (Mg doped) from the top to the bottom, respectively.

The presented diagrams prove that they represent all three principal cases: n-type, semi-insulating and p-type crystal, as indicated by position of the Fermi level close to conduction band, deep in the bandgap and close to the valence band respectively. Note that despite the band slope is small, nevertheless undergoes some change which indicate on the surface-bulk charge transfer.

Thus the above simulations provide verification of the adsorption properties of differently doped semiconductors. As it is shown, the adsorption energy of the adsorbed atoms does not depend neither on the field close to the slab (or equivalently on the excess electric charge on the surface state) nor on the Fermi energy which is determined by different doping in the bulk.

### 4. The Fermi level free surfaces

The Fermi level pinning is energetically costly, as the phenomenon entails charge separation in the layer close to the surface that is needed to create the electric potential variation, necessary to accommodate the bands so that the energy of the pinning state is at the Fermi energy in the bulk.

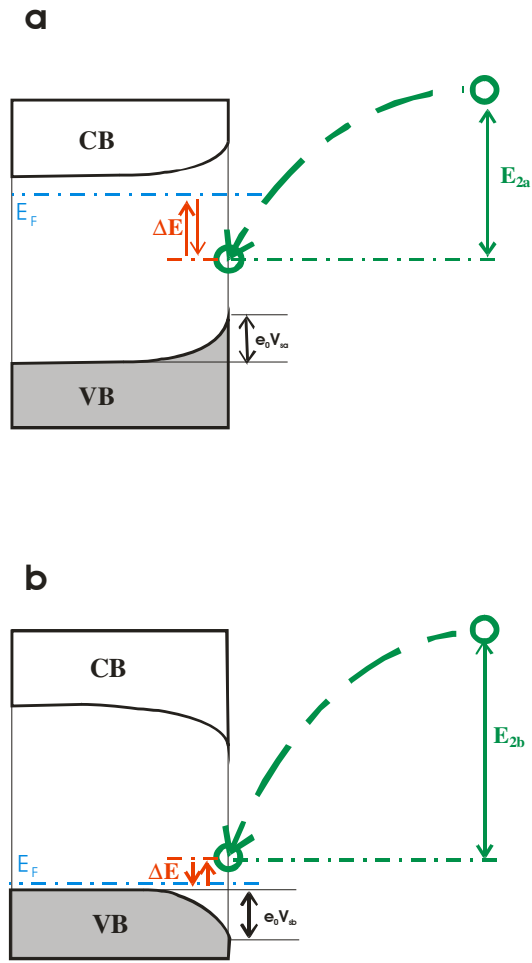

Fig. 7. The energy of the quantum states in the absolute scale, shown as changed during adsorption process at surfaces with Fermi level free at the surface having: a - surface acceptor, b - surface donor. The change of the energy of electron quantum state of the adsorbing species are denoted by green color, the energy change during charge transfer to or from the states in the bulk, close to Fermi level by red color.

As shown in Fig. 7, the absolute energy of the quantum states of the species adsorbed surface are not related directly to the doping in the bulk. Naturally as explained above, in the process without surface-bulk charge transfer the difference related to electrostatic potential at the surface cancels out. The surface-bulk charge transfer causes the energy contribution related to the difference of the Fermi and the surface state energy. As they could be of any value for nonpinned case for semi-insulating, n- and p-type bulk, this should be reflected in huge adsorption energy difference, observed for the various doping cases as shown in Fig. 8.

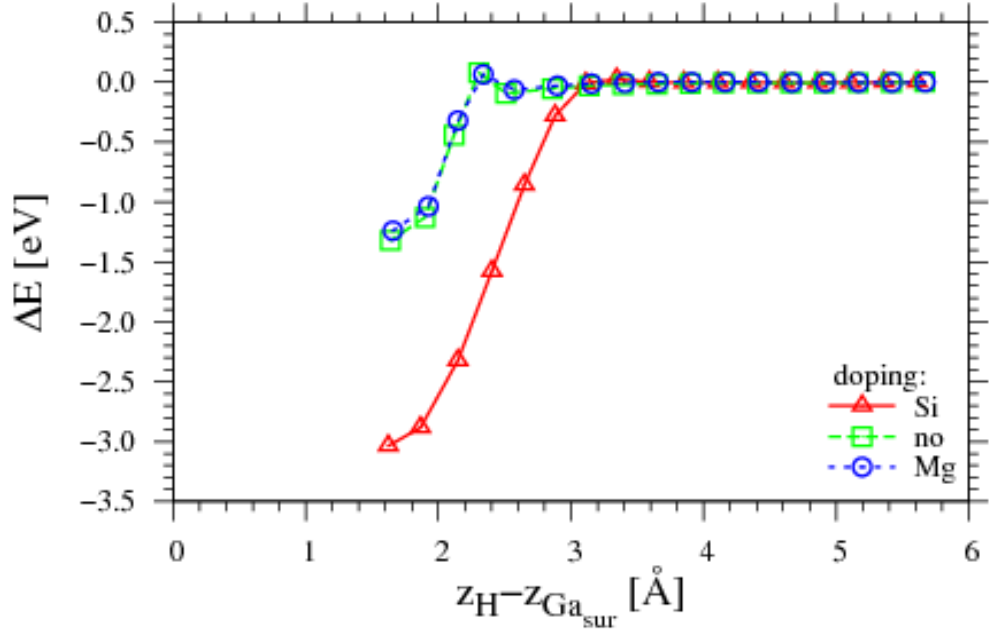

Fig. 8. Energy of the system during adsorption of single hydrogen atom on 8 double atomic layers thick 4 x 4 GaN slab, terminated by pseudo-hydrogen atoms with atomic number Z=0.75, representing n-type (Si doped), neutral (undoped) and p-type (Mg doped) GaN with (0001) surface covered with 0.75 ML of hydrogen (SIESTA). Effectively 12 sites have the hydrogen adatoms attached while the remaining 4 were uncovered.

The results of DFT calculation presented in Fig. 8 prove that the adsorption energy of hydrogen atom depends critically on the Fermi energy, i.e. on the doping in the bulk. This, entirely new result which has not been reported so far, was obtained by the application of the new technique capable to change both the field at the surface and the Fermi level in the bulk. In order to analyze this behavior, the band diagram and projected DOS, before and after adsorption of single H atom at 4 x 4 slab representing 0.75 ML covered GaN(0001) surface, i.e. for 12 sites covered by hydrogen with 4 remaining empty is shown in Fig. 9.

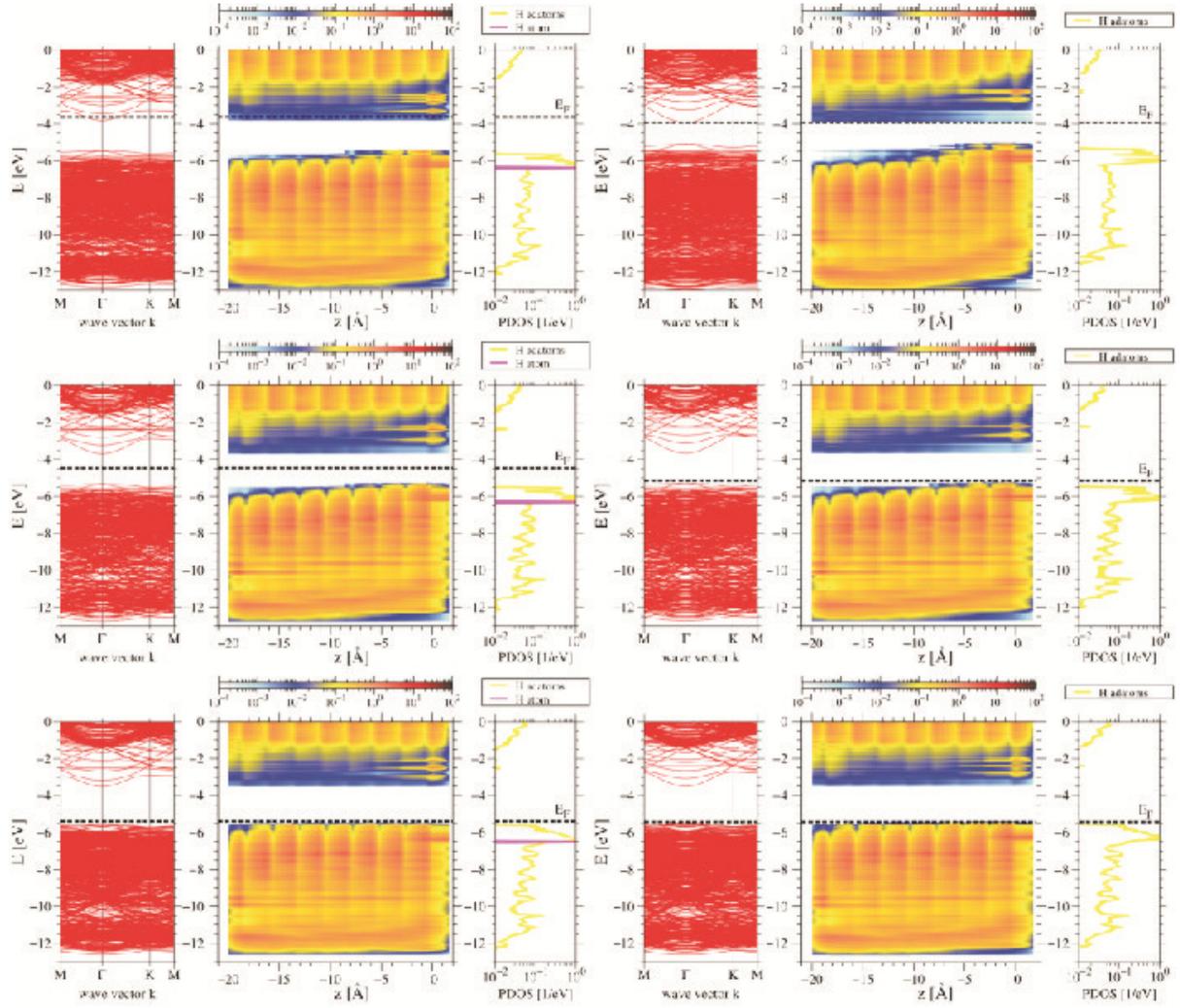

Fig. 9. Band diagram, atom row projected density of states and density of states projected on hydrogen atom, of the 4 x 4 slab before (left) and after (right) adsorption of single hydrogen atom at the 0.75 ML hydrogen covered GaN(0001) surface. The three cases present n-type (Si doped), neutral (undoped) and p-type (Mg doped) from the top to the bottom, respectively.

In these calculations, the termination atoms were selected so that the field within the slab is minimal. As it is shown in Fig. 9, by substitution doping method the Fermi level was shifted from the bottom of the conduction band (top), to the bandgap (middle) and finally to the top of the valence band. That corresponds to the n-type, semi-insulating and p-type material, respectively. As it is shown the slope of the band diagram increased after adsorption indicating that some electronic charge is shifted to the surface, i.e. surface is more of acceptor type. It is also worth to note that the energy difference of the Fermi level and the state of hydrogen adsorbed at the surface is identical for the semi-insulating and p-type. These two cases are drastically different from the n-type result where the difference is much higher. Therefore transition of electron from the Fermi level in the latter case creates much larger energy gain which is reflected by higher adsorption energy for Si-doped case shown in Fig 9.

The same behavior may be obtained using recently implemented simulation of charged point defects in SIESTA.[15] In this method uniform background charge is added, equivalent o some uniform band charge obtained by ionization of the point defect. The excess contribution represents the band charge, so it is smeared uniformly in the space covered by atomic orbitals. The appropriate addition simulated the absence and both doping types as shown in Fig. 10.

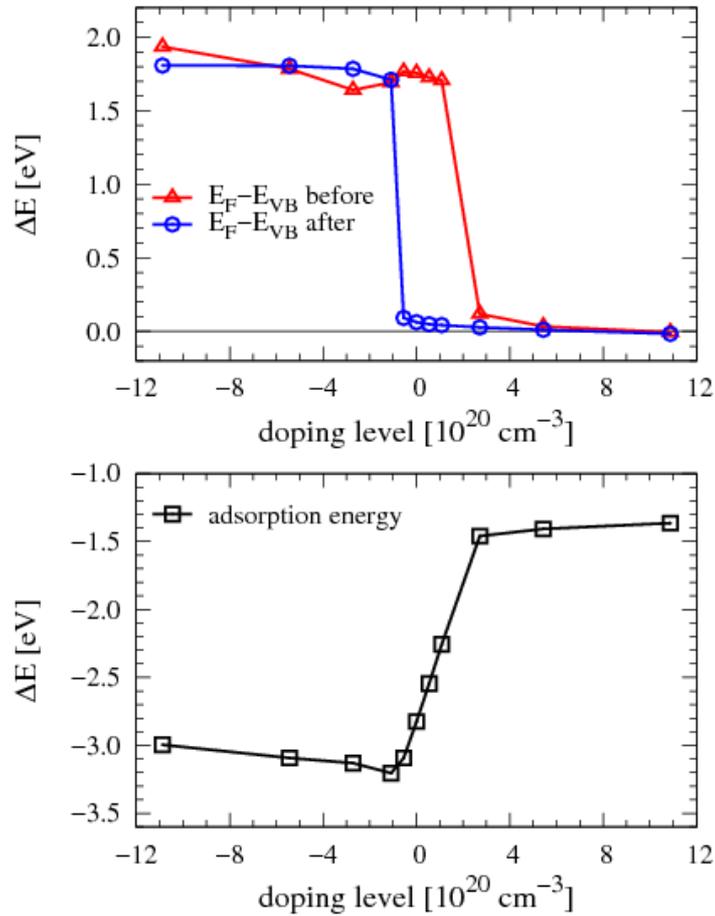

Fig. 10. Top - the position of the Fermi level with respect to the top of valence band, the bottom - the adsorption energy calculated as the difference of the energies of the 4 x 4 GaN slab covered by 11 (before) and 12 (after) hydrogen adatoms. For consistence the DFT energy of separate hydrogen atom was added to the energy of the slab covered by 11 hydrogen adatoms.

As it is shown change of doping covers the shift of the Fermi energy from degenerate hole to electron gas. Naturally, the transition is shifted because hydrogen adatoms are charged thus absorbing some of the charge from the bulk. The high positive and negative charge background correspond to the degenerate case, with small variation of the Fermi level only. The finite size of the system is visible in finite variation of the Fermi level data points.

The simulated cases correspond to nonpinned Fermi level at the surface as shown in Fig. 11. They prove again that the Fermi change affects drastically the adsorption energy of the system. It is shown also that the Fermi energy change is directly responsible for the change of the adsorption energy. For the small variation of the Fermi energy in both degenerate cases, both electron and holes, the adsorption energy is very small while the interval of drastic change of Fermi level corresponds to steep variation of the adsorption energy.

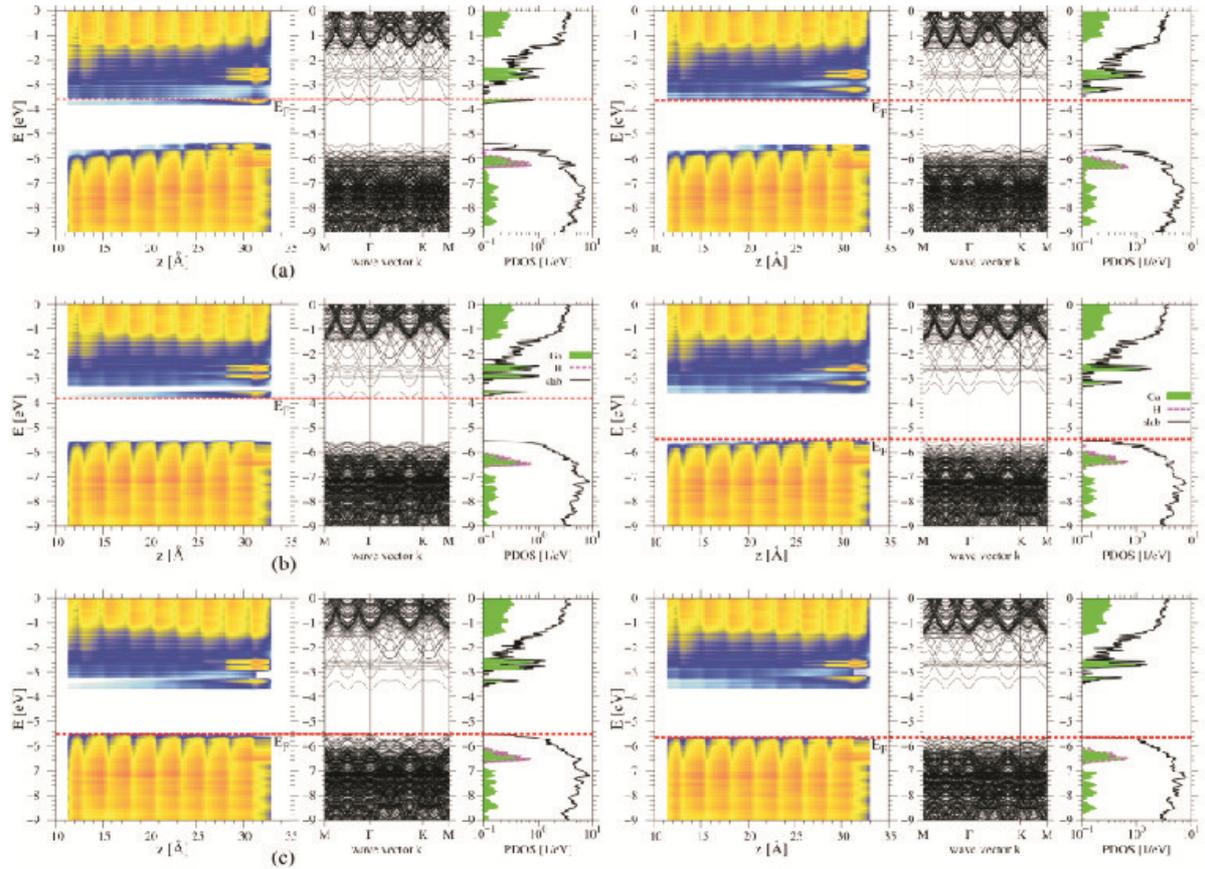

Fig. 11. Band diagram, atom row projected density of states and density of states projected on hydrogen atom, of the 4 x 4 slab, terminated by pseudo-hydrogen atoms with atomic number Z=0.735, representing GaN(0001) surface covered with 11 (left) and 12 (right) hydrogen adatoms. The three cases present n-type (n = 5.4 $10^{20}$ cm$^{-3}$), neutral (n = p = 0) and p-type (p = 5.4 $10^{20}$ cm$^{-3}$) from the top to the bottom, respectively.

The data confirm that SIESTA method is capable to change the system properties to simulate both n- and p-type doping. It is worth to note that the semi-insulating system is difficult to obtain, the Fermi level is close to valence or conduction band.

## 5. Summary

The adsorption of atomic and molecular species at the semiconductor surfaces with pinned and nonpinned Femi level was analyzed. The adsorption energy is determined by two contributions: the change of the energy of the electron quantum state of the species due to bonding and the charge transfer between the surface states and the bulk.

The first contribution is affected by the electron field at the surface shifting the energy of the surface states. Since the electron and nuclei are located in the same electric potential, the electrostatic contribution to the adsorption energy cancels out. Thus this contribution is independent of the Fermi level position, i.e. the doping in the bulk.

For the Fermi level pinned, difference of the energies of the surface states: the pinning one and the adsorbed species, is independent on the doping. Thus the charge transfer contribution is Fermi level independent, therefore the adsorption energy is independent on the doping in the bulk. The DFT simulations of adsorption of atomic hydrogen at clean GaN(0001) surface and atomic silicon at clean SiC(0001) surface confirm that the adsorption energy is independent on the position of Fermi level in the bulk. The small, of order of 0.1 eV change of the adsorption energy is caused by influence of the electric field shifting the quantum wavefunction of the surface state.

For the Fermi level nonpinned, the charge transfer occurs between the Fermi level in the bulk and the adsorbed states in the bulk. Thus the energy difference may be affected the by the Fermi energy change, i.e. is dependent on the doping in the bulk. DFT simulations of the adsorption of atomic hydrogen at 0.75 ML hydrogen covered surface demonstrate that the change of doping from n-type to p-type GaN changes the adsorption energy by about 2 eV. Thus large scale effect was confirmed by DFT calculations.

In summary adsorption energies are either Fermi level independent or sensitive for pinned and nonpinned semiconductor surfaces, respectively.

**ACKNOWLEDGEMENTS**

The calculations reported in this paper were performed using the computing facilities of the Interdisciplinary Centre for Modelling (ICM) of Warsaw University. This research was supported in part by PL-Grid Infrastructure. The research published in this paper was supported by funds of Poland's National Science Centre allocated by the decision no DEC-2011/01/N/ST3/04382.